%% file: main.tex
\documentclass{sigplanconf}[10pt]

\usepackage{graphicx}
\usepackage{eucal}
\usepackage{algorithm2e}
\usepackage{multirow}
\usepackage{amsmath}
\usepackage{amssymb}
\usepackage{theorem}
\usepackage{epsfig}
\usepackage{url}
\usepackage{hyperref}
\usepackage{multirow}
\usepackage{wrapfig}
\usepackage{comment}
\usepackage{color}
\usepackage{array}
\usepackage{ragged2e}
\newcolumntype{P}[1]{>{\RaggedRight\hspace{0pt}}p{#1}}
\newcolumntype{C}[1]{>{\centering\arraybackslash}m{#1}}

\numberwithin{equation}{section}	

\theoremstyle{plain}			

\theoremstyle{definition}		

\newcommand{\resistant}{\textsc{attack-resistant}\xspace}
\newcommand{\resistance}{\textsc{attack-resistance}\xspace}

\newtheorem{definition}{Definition}

\newcommand{\todobox}[3]{%
    \colorbox{#1}{\textcolor{white}{\sffamily\bfseries\scriptsize #2}}%
        #3 %
    \textcolor{#1}{$\blacktriangleleft$}%
}

\newcommand{\done}[1]{\todobox{blue}{DONE}{#1}}

\renewcommand{\done}[1]{}

\permission{}

\begin{document}

\makeatletter
\def\@copyrightspace{\relax}
\makeatother
	
\title{The Meaning of Attack-Resistant Systems}

\authorinfo{Vijay Ganesh}
           {University of Waterloo, Canada}	
		   {vganesh@uwaterloo.ca}

\authorinfo{Sebastian Banescu }
           {Technical University of Munich, Germany}
		   {banescu@in.tum.de}
		   
\authorinfo{Mart\'in Ochoa }
{Technical University of Munich, Germany}
{ochoa@in.tum.de}

\maketitle

\input{abstract}


\section{Introduction}

In the last several decades there has been impressive progress in the
development and application of automated software
testing~\cite{king1976sea,dart,CadarGPDE2006,SenMA2005} and formal
verification techniques~\cite{floyd,floyd71,DBLP:journals/cacm/Hoare69,clarke}. 
Despite these gains, establishing trustworthiness of software systems remains a 
notoriously difficult and expensive problem. Scalably guaranteeing 100\%
compliance of large software systems w.r.t rich specifications remains an open 
challenge. In fact, users simply assume that software systems will never be 
completely free of security vulnerabilities, despite our best efforts.

In response to this seeming inevitability of exploitable vulnerabilities in
software, security researchers have proposed ingenious defense
mechanisms (e.g., address space layout randomization abbreviated as
ASLR~\cite{aslr}, randomization for a variety of memory protection
schemes~\cite{diehard}, obfuscations~\cite{variaphd}, program
shepherding~\cite{kiriansky}, use of cryptographic
devices~\cite{sustek2011hardware}, and encrypted program 
execution~\cite{cohen1993operating}) that, under suitable assumptions, make it
``practically difficult'' for resource-bounded attackers to exploit
certain class of vulnerabilities even though vulnerabilities such
as buffer overflows are present in the system-under-attack. 
Many of these mechanisms are widely used in
practice because of their effectiveness against
common attackers. These defense mechanisms
are designed to complement traditional approaches such as testing and
formal verification, and provide an additional crucial layer of
security. 

On the other hand, in the past years several attacks have been
published against implementation of many of those defense mechanisms:
some of these attacks leverage side-channels (i.e. timing
side-channels against ASLR~\cite{shacham2004effectiveness} and
ISR~\cite{sovarel2005s}), while others challenge the assumptions made
on the attacker's strategy (i.e. by using ROP or JOP strategies
instead of straightforward injections \cite{rop,bletsch2011jump}).

Defense mechanisms distinguish themselves from formal methods and
testing in the following ways: First, they give an ``intuitive'' yet
practically effective {\it partial guarantee} of compliance of a software system
w.r.t a non-exploitability specification, even though the system may
have vulnerabilities in it. By {\it partial} we mean that a resource-bounded
attacker can still launch an attack but the probability is vanishingly
small, under suitable assumptions. By contrast, the traditional aim of
formal verification is a complete guarantee or establish the absence
of vulnerabilities. Testing techniques, on the other hand, are typically
incomplete and provide {\it no guarantees} of compliance of a large
software system w.r.t a specification. Second, defense mechanisms
offer different trade-offs between completeness and scalability, when
compared to testing or formal verification. For example, appropriate
use of defense mechanisms may lower the burden of establishing
security guarantees on software developers by not requiring them to
establish full compliance while giving some measure of security, at
the cost of possibly allowing certain hard-to-exploit vulnerabilities.

While the inevitability of vulnerabilities in software and the need for
effective defense mechanisms has been forcefully made by many
researchers in the past, and eloquently captured in a recent
paper~\cite{berger}, what seems largely missing is a formal
characterization of the security provided by these defenses, and to some extent,
a proof strategy or schema for establishing guarantees. Such a
formal and precise characterization of security, in terms of a partial
compliance guarantee, can be a foundational way of reasoning about and
comparing defense mechanisms of all kinds using the same standard
approach, just as in the field of cryptography, notions such as
``semantic security'' are standard ways of comparing encryption
schemes.

Although in cryptography attackers are usually seen as arbitrary
probabilistic algorithms with a certain number of resources (for
instance, number of gates which is polynomial in a security
parameter), we believe that in the case of vulnerability exploitation
and defense this is currently not possible: new attacker models often
radically challenge the assumptions made by defenders, as illustrated
by ROP, Blind ROP and JOP attacks \cite{bittau2014hacking}.  However,
we believe that making those assumptions explicit, and carefully
arguing why certain defense mechanism are secure against explicit
attacker classes would substantially improve the efficacy of defense
mechanisms, allowing defenders to spot subtle implementations or
design errors.

In this paper, building on paradigms from cryptography and program
analysis, we propose a general and formal definition of partial
compliance, we call \resistance, of a software system w.r.t a
(non-exploitability) specification. The key insight of our formal
program analysis approach is that a program $p \in P$ together with
its defense mechanisms $d \in D$ (written $p+d$) is \resistant against
a well-defined attacker (or attack model) $\mathcal{A}$ if the code
and runtime behavior of $p+d$ is resistant to exploitation by
$\mathcal{A}$, and this can be expressed using the
complexity-theoretic idea of {\it negligible success probability} from
cryptography. Informally, we say that $p+d$ is \resistant to attacker
$\mathcal{A}$ if the probability of successfully launching a remote
program execution by the attacker is negligible in a suitable security
parameter. Based on this definition, we propose a formal analysis
framework and a research program whose goal is to analyze the efficacy
of defense mechanisms from a variety of contexts. We also discuss
building blocks for a proof strategy that can be useful to show
\resistance in various contexts, and that consists in reducing the
attack probability to the likelihood of guessing a random seed, which
lies at the heart of most defense mechanisms, in analogy to
Kerckhoffs's principle in
cryptography~\cite{kerckhoffs1883cryptographie}.

In sum, we make the following contributions in this ``Vision and
Challenges'' paper:

\subsection*{Contributions}

\begin{enumerate}

\item We provide a formal definition of \resistance and discuss a
  high-level proof strategy for it.

\item We show the utility of this definition and the proof strategy by
  analyzing and comparing different ISR implementations from the
  literature.

\item Finally, we suggest that unlike formal verification of arbitrary
  programs to establish their non-exploitability (e.g., absence of
  overflow vulnerabilities), in some cases it is possible to show
  \resistance for a large class of programs by only paying the cost of
  formally verifying the defense mechanism exactly once.
\end{enumerate}

Note that our goal is the quantification of the effectiveness of
probabilistic defenses in a general sense, although in this paper we
focus primarily on defenses for memory safety vulnerability
exploitation. On the one hand, we do not aim to cover all types of
run-time countermeasures to memory safety vulnerabilities, but only
the ones in the probabilistic category as defined by Younan et
al. \cite{younan2012runtime}. On the other hand, our approach can also
be used to analyze probabilistic defense mechanisms beyond memory
safety, such as SQLrand \cite{boyd2004sqlrand}, which we plan to do as
part of future work.

\section{Attacker Model}

The adversary or attacker $\mathcal{A}$ is defined as a probabilistic
algorithm that has access to a copy of the source code of the
program-under-attack $p \in P$ (e.g., a web server) and is aware of
all low-level vulnerabilities contained by $p$, such as buffer
overflows, format string vulnerabilities etc. The assumption that the
attacker has the source code of $p$ is realistic, given that binary
code and obfuscations that are not cryptographically-secure can be
easily reverse-engineered by attackers. The attacker is allowed to run
any type of program analysis (static and/or dynamic) within a time
bounded by a polynomial over the length of a security parameter (a
random \emph{seed}). The attacker's goal is to learn the
vulnerabilities in the program $p$, construct and launch appropriate
exploits.

$\mathcal{A}$ has remote access to a machine $\mathcal{M}$ on which
the vulnerable program $p$ is running (e.g., a machine running the
Apache Web Server). $\mathcal{A}$ can interact with $p$ via a network
channel, where it can send multiple messages and observe their
respective output. More precisely $\mathcal{A}$ can produce $k$
messages: $m_1,m_2, \dots, m_k$, where $m_i \in I$ and $I$ is an input
message space.  Each input message is represented by finite bit
strings, possibly containing exploits and payloads.  $\mathcal{A}$
receives $k$ outputs from $p$, one for each input message: $o_1,o_2,
\dots, o_k$, with $o_i \in O$, where $O$ is the output message space.
The adversary encodes a malicious program $p'$ in each input message
using a probabilistic function that adaptively takes into account
previously received outputs (at most $k$) $e: P \times O^{k} \to I$.
The attacker is not allowed side-channel access to the runtime of $p$
(beyond what is leaked through its observations) or physical access to
$\mathcal{M}$. Note that although we leave $\mathcal{A}$ largely
unspecified to account for adaptation to the interaction with $p$, we
assume that some characteristics of the encoding function $e$ are
known (for instance that is tailored for ROP exploit construction,
shellcode injection etc.). In the following we will thus denote the
adversary as $\mathcal{A}_e$.

Note that if this function is completely unknown, it will be practically 
impossible to show \resistance of a particular defense mechanism, since  
new attack techniques can easily invalidate certain defense assumptions (as
shown by ROP and recently by blind ROP \cite{bittau2014hacking}).

If $p$ contains a buffer overflow vulnerability, it allows
uploading the payload into a part of $p$'s process memory. 
If the exploit is successful it leads the control flow of $p$ to 
the payload causing it to execute, thus resulting in a control-hijack
attack (for further reading, we refer to the rich literature already
out there on buffer overflow attacks in C/C++
programs~\cite{erickson2008hacking,rop}).

Formally, the attacker $\mathcal{A}_e$ wants to remotely execute a malicious 
program of his choice $p' \in S \subset P$ on a victim machine 
$\mathcal{M}$. We define $P$ to be the set of all programs expressed
as machine code runnable on $\mathcal{M}$ (for instance x86 machine code), 
and $S$ is a subset of $P$, containing malicious programs that are the target
of the attacker and not containing $p$. We leave $S$ unspecified and strictly 
smaller than $P$ for two reasons: on the one hand, it does not make sense that 
$S$ contains $p$, since this program is being executed on $\mathcal{M}$
anyway, on the other hand the notion of \emph{malicious} can vary from 
context to context (there are many programs that are harmless if injected).

To more precisely define the goal of the attacker we consider the following
event:
$$ E(p, p',\mathcal{A}_e ) = \{ \mbox{ $p'$ runs in process $p$ after interaction 
with $\mathcal{A}_e$ } \} $$
By $p'$ runs in process $p$ we mean that there is a sequence 
of assembly instructions executed in the process memory of $p$ that is 
semantically equivalent to $p'$. Note that without loss of generality we can 
assume that for all $p' \in S$, $S$ contains all programs semantically 
equivalent to $p'$. 

\section{Attack Resistance}

In this Section we define the notion of \resistance more formally.

\begin{definition}
We say a defense mechanism $d(n) \in D, \ n \in \mathbb{N}$ provides attack resistance for a class of victim program $\mathcal{C}$ running on a machine 
$\mathcal{M}$ against an attacker $\mathcal{A}_e$ and a set of malicious 
programs $S$ if and only if:
$$\forall \ p \in \mathcal{C} \ \forall \ p' \in S  \ \Pr[ E(p+d, 
p',\mathcal{A}_e)] \leq \epsilon(n),$$
where $\epsilon$ is a negligible function on $n$.

\done{SB: Probably no defense mechanism alone is \resistance. However, their combination is. Even in the example we have for the HotSpot paper we needed PointGuard together with a monitor that prevents leakage of the key. Those are 2 defense mechanisms not one. Should we adapt the definition and the following examples?}

\end{definition}

Clearly $\mathcal{C}$ excludes program whose vulnerability free version 
allow users to execute programs in $S$ (for instance interpreters). However
we leave this class of programs unspecified, since in general it will depend on 
the defense mechanism and the attacker model considered.

We give a few intuitive examples here and a detailed example in Section~\ref{sec:example}:
\begin{itemize}
	\item If $e$ is simple code injection (appending malicious code to
	a message that overwrites a buffer) then some implementantions of ISR 
    \cite{kc2003countering} provide attack
	resistance for programs $p$ that do not leak the seed. We will elaborate
	about this example in the following section.
	
	\item If $e$ is ROP, then simple ASLR does not provide attack resistance
	for many programs,
	whereas advanced ASLR (where the base address of $p$ is also 
	randomized) or advanced ISR \cite{papadogiannakis2013asist} do.

	\item If $e$ is blind ROP \cite{bittau2014hacking} and $k$ is small 
	(or it is not allowed to restart the program enough times by the same 
	attacker), then common ASLR provides attack resistance against
	a large class of programs, otherwise it does not.
	
\end{itemize}


\subsection{Proof Strategy for Establishing Attack Resistance} \label{sec:proof-strategy}

Note that in practice most defense mechanisms against memory
management exploitability rely on the randomness of a certain run-time
transformation. Ideally this randomness should guarantee security and
not the complexity of the transformations induced by the defense
mechanisms, similarly as secret keys are the security cornerstone of
cryptographic guarantees as suggested by Kerckhoffs's
principle~\cite{kerckhoffs1883cryptographie} or Shannon's
maxim~\cite{shannon1949communication}.

Therefore, a strategy for establishing \resistance is to prove that
the following sufficient conditions hold for a program $p \in
\mathcal{C}$ and the defense mechanism $d$, under the attacker model
$\mathcal{A}$:

\begin{enumerate}

\item {\bf Runtime Secret Seed:} A condition for the efficacy of
  defense mechanisms is the existence of bit sequence $s, |s| \in
  O(n)$ called {\it Runtime Secret Seed} in the pair $p+d$ of program
  and defense mechanism. Note that if the transformation induced by
  the defense mechanism is not probabilistic, then an attacker could
  eventually reverse engineer it.

\item {\bf Resistance guaranteed by keeping the seed secret: } A
  further condition is a Lemma relating the attack event (remotely
  executing a malicious program) with the probability of guessing the
  random seed $s$.  In other words, to show that for each $p' \in
  S$, $$\Pr[E(p+d, p',\mathcal{A}_e)] \leq \alpha \cdot \Pr[\{
    \mathcal{A}_e \mbox{ guesses $s$} \}]$$ for some constant
  $\alpha$.

\item {\bf Bounding the Attacker's probability of guessing $s$:} A
  useful formal notion to measure the attacker's knowledge is to
  consider the entropy of $s$. A further condition for the efficacy of
  the defense mechanism $d$ is thus that the attacker's a priori
  knowledge of $s$ must be low, i.e., the a priori entropy of $s$ is
  high, and that it remains high after static and dynamic analysis of
  $p+d$.  This notion can be formally captured for instance by
  quantifying maximal leakage \cite{smith2009foundations}, seen as the
  difference between the a priori and posteriori entropy of $s$ after
  program analysis and interaction with the victim machine.
  This condition is important since weak defense mechanisms could have a poor
  key generation algorithm or some other vulnerability that decreases the
  ideal (maximal) entropy of the key.
  In other words to show is that the 
  $\Pr[\{ 
  \mathcal{A}_e \mbox{ guesses $s$} \}] \leq \epsilon(n)$
	for instance by computing the maximal 
	leakage of $s$ on the network channel (defined as the difference of the 
	entropy of $s$ before executing $p$ and after code analysis and 	observing the $k$ output messages).


\end{enumerate}


In the following section we will discuss some variants of ISR from the point
of view of \resistance.

\section{Attack-resistance of ISR} \label{sec:example}



In the following we discuss and analyze ISR as proposed in \cite{isetrandomization}, which is supposed to be resistant against \emph{code-injection attacks}. 
In \emph{code-injection attacks} a sequence of machine code instructions are written to the stack- or heap-memory of a running process by exploiting a buffer overflow vulnerability.
Additionally, the attacker must overwrite the \emph{return address} on the current stack frame, using a value that points to the beginning of the injected code.
After the current function ends its execution, the overwritten return address is loaded in the \emph{instruction pointer} register and starts executing the injected code.

\begin{figure}[h!]
  \centering
    \includegraphics[width=0.4\textwidth]{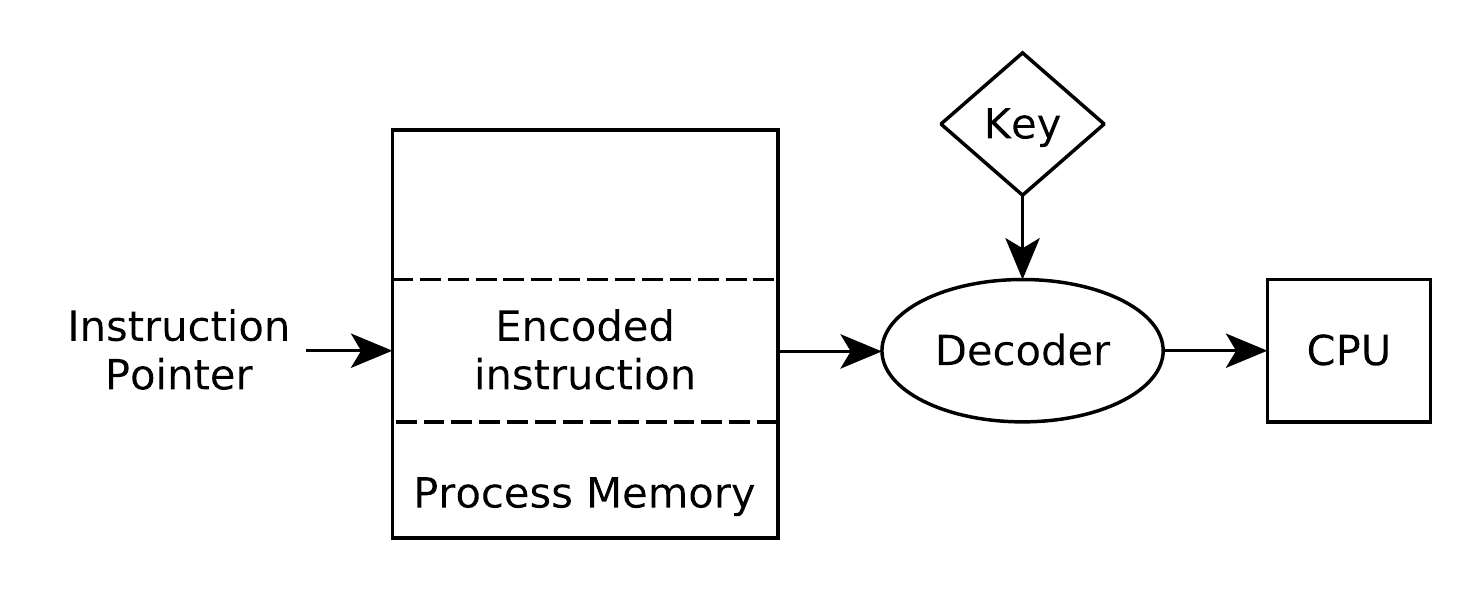}
    \caption{ISR run-time instruction decoding before execution.}
  \label{fig:isr}
  
\end{figure}

The idea behind ISR (illustrated in Figure~\ref{fig:isr}) is to encode the original instructions of a program using a random key and decode them during runtime, right before execution of each instruction.
Both RISC (e.g.~ARM) and CISC (e.g.~Intel \emph{x86}) architectures use machine code instructions consisting of \emph{opcodes} followed by zero or more \emph{operands}.
Encoding using a random key maps the values of opcode and operands to a set unknown to a remote attacker which does not have access to the key.
Therefore, if the attacker does not guess the correct key, the injected code will be ``decoded'' before execution to a different value than intended. 
Depending on the number of possible instructions for that architecture, their length and the length of the key, the program will either crash or execute a different instruction than intended. 
We will argue about the probability of each of these outcomes in the following paragraphs. 
However, since there have been several attacks on ISR~\cite{sovarel2005s, weiss2006known}, which are enabled by the existence of 1 and 2 byte instructions on CISC architectures such as Intel \emph{x86}, we will assume only 32-bit RISC architectures in the following paragraphs.
Moreover, Table~\ref{tab:attacks} shows a broader overview of the previously mentioned attacks (first row) and the assumptions they rely on (first column). 
We give a few countermeasures on the right-most column which disable the attacks.
In the following we will use a combination of some of these countermeasures to build a defense which is \resistance against code-injection attacks.

\begin{table*}
\begin{center}
\begin{tabular}{|P{5cm}||*{6}{C{1cm}|}|P{3cm}|}
\hline
Assumptions (down) \textbackslash Attacks (right) & Return attack \cite{sovarel2005s} & Jump attack \cite{sovarel2005s} & Direct key extraction \cite{weiss2006known} & Known plaintext attack \cite{weiss2006known} & Chosen key attack \cite{weiss2006known} & Key guessing attack \cite{weiss2006known} & Countermeasures\\
\hline\hline
Attacker knows address where code is inserted & $\times$ & $\times$ & × & × & × & × & ASLR\\
\hline
Same key used after crash and restart & $\times$ & $\times$ &   &  & × & $\times$ & Change key upon crash\\
\hline
Encoded instruction and decoded instruction pair give the key (e.g. XOR is used for de-/en-coding) & $\times$ & $\times$ & × & $\times$ & × & $\times$ & Use bit transposition\\
\hline
Attacker can distinguish 2 behaviors (e.g. crashed and not crashed) & $\times$ & $\times$ & × & × & × & $\times$ & ×\\
\hline
Attacker can distinguish server activity between guessed return instruction and first instruction that would cause the program to crash & $\times$ & × & × & × & × & × & \\
\hline
Attacker can produces infinite loops on the target program & × & $\times$ & × & × & × & × & \\
\hline
Key location is fixed and in program accessible user-space memory & × & × & $\times$ & × & × & × & Make key location unaccessible via user-space program\\
\hline
Format string or buffer overflow vulnerability that displays string to end-user & × & × & $\times$ & $\times$ & × & × & Secure Multi-Execution\\
\hline
Knowledge of the binary executable code (this is the plaintext). It could also be a library if that library is ISR-ed & × & × & × & $\times$ & × & × & Secure Multi-Execution\\
\hline
Key(-generator) of ISR mechanism can be replaced via remote attack (e.g. ret-to-libc) & × & × & × & × & $\times$ & × & Make key location unaccessible via user-space program\\
\hline
\end{tabular}
\end{center}
\caption{ISR Attacks versus Assumptions and Countermeasures}
\label{tab:attacks}
\end{table*}

\paragraph{ISR with XOR-ing in ECB mode:}

Consider a hardware implementation of ISR using a fixed length $n$-bit key (stored per process in kernel space), which changes on each process start-up.
Furthermore, assume that the \emph{Decoder} in Figure~\ref{fig:isr} is a $n$-bit XOR gate implemented in hardware.
Let $\mathcal{IS}$ the instruction set for a given RISC architecture. 
Let $|\mathcal{IS}| = 2^l$. Let $n$ be the size in bits of an instruction.
Now assume that adversary encodes a program $p'$ using $e$. Irrespective
of his encoding strategy, the first instruction $i_1$ of $p'$ must
be such that $i_1 \in \mathcal{IS}$, otherwise the program will crash.
If the key $s$ is chosen uniformly at random, and has not been leaked:

$$\Pr[\{i_1 \in \mathcal{IS}\}] \leq \frac{1}{2^{n-l}}.$$

because XORing with a random value is basically a random permutation in the set of possible instructions. Since $l$ is costant for a given architecture, this function is negligible in $n$. \done{SB: but $n$ is also a constant here because it doesn't make sense to increase the size of the key more than the length of the instructions. Does it make sense then to say that ``this function is negligible in $n$''?}

For example, the 32-bit ARM architecture has an instruction set consisting of under $2^{7}$ instructions. 
Therefore in this case we obtain that the probability that an injected instruction is in $\mathcal{IS}$ is less or equal to $\frac{1}{2^{25}}$. 
Otherwise, the program will crash and will use a different key upon restart,
provided there is no leakage on the key.

\paragraph{Preventing leakage:} In the event that the program does not crash and the key is wrong, the program will execute some instruction from $\mathcal{IS}$.
To avoid that this information or another vulnerability in the program leaks the key such as in \emph{direct key extraction} and \emph{known-plaintext key extraction} presented in \cite{weiss2006known}, we propose
to combine ISR with \emph{secure multi-execution (SME)}~\cite{devriese2010noninterference}. 
In this case, we SME to run two copies of the program in parallel, each randomized with a different key.
Both programs receive the same input, however, before outputting anything the SME mechanism checks if the outputs of the two program copies are equal.
If so, it forwards the output, otherwise the output is blocked.
Under the assumption that the behavior of the program is deterministic, SME guarantees that any of the two keys will not be leaked. Thus,
after proving absence of leakage of the key, we can conclude:

$$\Pr[E(p+d,p',\mathcal{A}_e)] \leq \Pr[\{i_1 \in \mathcal{IS}\}] \leq \frac{1}{2^{n-l}} \leq \alpha \cdot \epsilon(n).$$

for $\alpha = {2^l}$ and $\epsilon(n) = \frac{1}{2^n}$.


\paragraph{ISR with a Stream-cipher:}

An alternative way to implement ISR is, instead of XOR-ing each instruction in the program with the same key, to use a \emph{pseudo-random number generator} (PRNG) with a large periodicity, seeded by our $n$-bit key in order to decode the programs with a stream-cipher. In this case, the length of an instruction
is independent of the size of the key, and therefore we denote it with $n'$.
If the malicious program $p$, which the adversary wants to inject consists of $j$ instructions $[i_1, i_2, \cdots, i_j]$, a successful attack requires either guessing $j \times n'$-bits or the $n$ bits of the key.
Therefore, the probability that the program does not crash and chooses another random key is:
$$\Pr[\{i_1, i_2, \cdots, i_j \in \mathcal{IS}\}] \leq \frac{1}{2^{min(j(n'-l),n)}},$$
which is far less than $\Pr[\{i_1 \in \mathcal{IS}\}]$.
The \emph{key-guessing attack} presented in \cite{weiss2006known} shows some examples of the shortest sequences of instructions that cause damage consist of at least 3 instructions. 
Therefore, if we consider the previous example of a 32-bit ARM architecture and substitute $j$ by 3, we get that the probability of executing a sequence of injected code consisting of 3 instructions without crashing the program is equal to $\frac{1}{2^{75}}$. Interestingly, in this case is \emph{not} sufficient to increase the key (seen as the input to the PRNG) size to achieve
asymptotic better guarantees, but it is necessary to also increase the size
of the instructions.


\paragraph{Increasing the instruction size:} Note that the effective security
parameter in ISR is the instruction size. Usually, the instruction size is 
determined by the CPU architecture, and therefore to increase it, one would
need tailored hardware. Although it might be  unpractical (from
the efficiency perspective)
to increase the instruction size to improve upon the security guarantees,
it is possible to achieve this without modifications in hardware by using
virtualization. For instance, virtualization obfuscation such as the one
provided by Tigress\footnote{\url{http://tigress.cs.arizona.edu/}} allows one to configure the size of the virtualized
and randomized instruction set. However note that as discussed before, in 
practice the probability
of code injections attacks for ISR on RISC (32 bit) based architectures are already vanishingly small, given guarantees on the absence of key leakage.

\paragraph{Issues with side channels:}
Under the assumptions that stack-base address randomization is not employed and that the key is not changed when restarting a program and Sovarel \emph{et al.}~\cite{sovarel2005s} use timing side-channels to determine the key value on Intel \emph{x86} architectures. Their technique has a success rate of over 95\% and requires less than 1 hour for 4096-byte keys. However, in the previous paragraphs we have assumed that the key changes on each program restart, which is enough to thwart this attack.

\paragraph{Partial leakage:}
Note that the fundamental goal to show \resistance is that the bounding 
probability function $\epsilon(n)$ is negligible. Therefore, it is possible
to allow some leakage on the key, if it is possible to show that this will
not break the negligibility of the bounding function. This is for instance 
the case when leakage is constant, independent of the key size, or when
it is logarithmic. In future work we plan to explore strategies to account
for the possibility of side-channels that have such properties. This is
important for efficiency, since for instance resampling a key after every crash
is computationally expensive. However, allowing a limited number of crashes 
with the same key will bound the information that an adversary can gain. 
For instance, leakage resilient cryptography has been sucessfully applied
to real implementations in the case of side-channels caused
by CPU caches \cite{barthe2014leakage}. Alternatively, bounds on the leakage
could be obtained by leveraging on quantitative estimations on implementations 
using techinques such as \cite{mccamant2008quantitative}.

\paragraph{Other attacker models:}
Note that so far we have focused on common injection attacks. As we have discussed before, ISR together with secure multi-execution, is not resistant against other attack strategies such as ROP  \cite{rop} attacks.
ROP attacks do not inject code in process memory, instead they inject addresses of gadgets, which are not affected by the simple ISR implementation so far.
However, to make the program \resistance against ROP, it is possible to extend 
ISR with defense mechanisms such as kBouncer~\cite{pappas2013transparent}, ASLR 
(randomizing the program base address), or other mechanisms that invalidate ROP 
(such as hardware implementations of ISR \cite{papadogiannakis2013asist}). 
However it is not trivial to show that the combination of this strategies
provide \resistance: they might use different seeds (of different sizes) to
achieve randomness on their own and they might introduce new side-channels
or attack vectors. Moreover reasoning about attacks such as ROP is more 
challenging, since it involves reasoning about the probability that a 
randomized injected address does not hit a gadget, which will strongly depend
on the particular layout of the memory at the time of attack. In the future
we plan to analyze this issues more carefully, and to show \resistance of
this mechanisms under certain assumptions.

\done{expand
on this point, given space.}

\section {Threats to Validity of the Research Program}

\paragraph{Practical Value of The Proposed Formal Analysis
  Framework, and its difference with respect to traditional formal
  verification:} The proposed analysis framework essentially
establishes that cryptographic primitives used by the defense
mechanism $d$ prevent exploitation by certain attackers, and that the runtime
information-flow leakage technique used by $d$ guarantees a bounded 
leakage of secret bits. This is a one time
cost of analyzing $d$ for a large class $\mathcal{C}$ of
programs that can
be accomplished by using appropriate formal verification
techniques, but leaves open that possibility that other attackers
can effectively exploit the program. By contrast, in traditional formal 
verification approaches, we have to prove that every program $p$ is completely 
free of memory safety vulnerabilities, if our goal is a 100\% guarantee
that $p$ adheres to a non-exploitability specification.

\paragraph{Model-checking is PSPACE-Hard:}
Model-checking a safety property is known to be PSPACE-Hard and so then
exploit-construction should be equally hard (in general undecidable)
from a complexity-theoretic point of view. In what way are the
guarantees being proposed here any stronger? Observe that we are not
saying anything about the general exploit-construction problem, and
instead focus on analysis of a parametric class of programs protected
by a defense mechanism. The PSPACE-Hardness result is general, and
doesn't say anything about the hardness of exploiting a given
program. We are  proposing that given a parametric class of
programs, certain defense mechanisms can make it harder to exploit and
this guarantee of hardness is captured precisely using the notion of
\resistance.

\paragraph{The Defense Mechanisms considered are Useful only for Memory Errors:}
The complex software ecosystem of today is subject to all manner of
attacks, not only buffer-overflow attacks. Hence, it pays to formally
analyze defense mechanisms and distinguish effective ones from the
rest.  The idea of \resistance can be used to analyze any defense
mechanism that relies on proper implementation and use of cryptography
and guarantees against remote execution of malicious code, such
as SQLrand \cite{boyd2004sqlrand}.

\paragraph{Some Defense Mechanisms Shift Attacks on Integrity and/or 
Confidentiality to Attacks on Availability:} We have assumed that $d$
detects and prevents all attacks against $p$. However, in practice it 
is often the case that an attack is ``detected'' by a crash of $p$, which 
prevents an attacker from taking over the control-flow of $p$. This is
the case for defense mechanisms such as PointGuard or its variant we present 
here. The threat to validity is that \resistance 
refers to integrity and/or confidentiality attacks, but not availability.

\paragraph{Guaranteeing the absence of leakage is hard:} Depending on the
precise details of a defense mechanism, it can be in general harder to
guarantee the absence of leakages (non-interference of the secret
random see w.r.t. public outputs) than show the absence of BOFs, since
non-interference is not a safety property, i.e., is a 1-hyper-safety
property. Moreover, there is the side-channel issue discussed before:
in general to guarantee the absence of certain side-channels, there is
a performance penalty to be paid (such as shutting down CPU caches,
constant time cryptography etc.).

\section {Related Work}

There is extensive use of formal methods in the security context,
especially for protocol analysis~\cite{datta2007protocol}. There is also
considerable prior work on dynamic information flow
leakage~\cite{schwartz2010all}, and on analysis of implementation of
cryptographic primitives~\cite{appelverification}. Our work is
different in that we propose the combination of such analyses to
formally establish the effectiveness of defense mechanisms with respect to
clearly specified attackers.

Pucella and Schneider~\cite{pucella2010independence} have also formally 
investigated 
the effectiveness of randomized defenses in the context of memory safety. Their 
main result is to characterize such defenses as to be probabilistically 
equivalent to a strong typing which would guarantee memory safety for buffers, 
thus reducing the security of
the defense mechanism to the strength of strong typing. In particular, they analyze \emph{address obfuscation}~\cite{bhatkar2003address}, a defense 
mechanism against memory corruption attacks, that uses a secret key to randomize 
the offsets of code and data in heap memory. Their idea is to treat address 
obfuscation as a probabilistic type checker, which has a certain probability $p$
of crashing the program when a buffer overflow occurs. Differently from our 
discussion on
the \resistance of ISR against code-injection attacks in 
Section~\ref{sec:example}, 
by relating to a type checker that catches out of bound access, they 
automatically consider a wider range of possible attacks. However, also due to 
this abstract charachterization, they acknowledge the difficulty on computing 
the probability of a successful attack, as we propose 
to do. We believe that different from~\cite{pucella2010independence}, 
the \resistance framework proposed in this work can also be used to compute the 
probability 
of a successful buffer overflow against address obfuscation if we combine this 
defense technique with other techniques, such as SME or quantitative 
information flow analysis, that prevent substantial leakage of the secret key.


Abadi and Plotkin~\cite{abadi2012protection} also formally investigate
the effectiveness of memory layout randomization in programming
language terms.  Specifically, they consider layout randomization as
part of the low-level implementation of a high-level language. This
implementation is analyzed by ``mapping low-level attacks against it
to context in the high-level programming
language"~\cite{abadi2012protection}. Their results are phrased as
full abstraction theorems that say that ``two programs are equivalent
in the high-level language if and only if their translations are
equivalent (in a probabilistic sense) in the low-level language. The
equivalences capture indistinguishability in the presence of an
arbitrary attacker, represented as the context of the
programs''~\cite{abadi2012protection}. These equivalences are powerful
enough to express both secrecy and integrity properties. Jagadeesan et
al.~\cite{jagadeesan2011local}, extend the work of Abadi and Plotkin
by considering return-to-libc attacks as well as dynamic memory
allocation features in programming languages. While these are very
powerful results, they are less general than our approach because we
consider a wider variety of defense mechanisms and attacks, and do not
fix the programming language. By contrast, the results of
\cite{abadi2012protection,jagadeesan2011local} are restricted to
ASLR. On the other hand, if a programmer were to write their programs
in the high-level language of
\cite{abadi2012protection,jagadeesan2011local}, then they would get
the protection of ASLR for free without having to explicitly prove
them. In our setting, someone has to prove for a given class of
programs, that the defense mechanism is effective for that
class. Agten et al. \cite{agten2012secure} describe a compiler that
implements full abstraction, i.e., the compiler guarantees that
standard security features in languages such as Java (e.g., private
field) are compiled correctly down to the lower-level language and the
security at the low-level is enforced via fine-grained access
control. This work is orthogonal to ours since, unlike our approach,
they don't analyze probabilistic defense mechanisms.

Sovarel et al.~\cite{sovarel2005s} and Weiss et
al.~\cite{weiss2006known} also perform an analysis of the attacks
against ISR presented in Table~\ref{tab:attacks}.  They compute the
probability of a successful code-injection attack on a program
protected by the ISR mechanism presented in
\cite{barrantes2005randomized}.  Differently from our analysis they
only give numeric values for these attacks on Intel \emph{x86}
architectures under additional assumptions, without offering a more
general overview of this probability in terms of the size of the key
and the size of the instruction set and instruction sizes.

Shacham et al.~\cite{shacham2004effectiveness} presented a study
regarding the assessment of randomization based software
defenses. Their study consisted of a concrete attack on any program
containing a buffer overflow vulnerability protected by
ASLR~\cite{aslr}. Their attacker model was similar to our in the sense
that the attacker had knowledge about the vulnerabilities in the
target program and only had remote access to the target
machine. However, one key factor in their attack was the assumption
that the remote machine is running a certain version of an Apache
server, whose behavior is used as a side-channel. More specifically,
if the correct offset of a libc function is sent to the Apache server,
then it has a different behavior than when the guess was incorrect. Of
course the attack is not dependent on Apache itself, but on a
side-channel which leaks the RSS of ASLR. This attack is highly
effective on 32-bit systems where, ASLR only randomizes the base
address of memory pages which are 4 KBs in size. Therefore, there are
only 20-bits of memory which can be randomized. Brute-forcing a
randomization space of this size is practical on current
hardware. Therefore, the straightforward solution to this attack is to
switch to a 64-bit system. In contrast to
\cite{shacham2004effectiveness} we assume that side-channels which
leak the RSS do not exist.

\section{Conclusion}

In conclusion, we have proposed a new notion of partial compliance
that we call \resistance using the notion of negligible success 
probability from cryptography. We have also shown that
if a well-defined attacker's uncertainty of a random seed $s$ remains
high after interaction with $p+d$, and the only way to attack the system
is by guessing $s$,
then we can say that $p+d$ is \resistant, and that the defense mechanism
$d$ is effective. We argue that this notion can be very helpful in
analyzing, and helping design new defense mechanisms. Furthermore,
such analysis can be useful in recognizing design and implementation
weaknesses of defense mechanisms that are not \resistant.
As part of future work we plan to formalize analyze the
\resistance of further popular defense mechanisms and
their combinations, to discuss their
strength against common attacker models, and to explore relaxations of
information flow guarantees (such as quantitative bounds on entropy
reduction), to account for  efficient implementations of defense
mechanisms. Moreover, we plan to explore extensions of our framework to cope
with probabilistic defense mechanisms beyond memory safety attacks, such
as SQLrand \cite{boyd2004sqlrand}.

\paragraph{Acknowledgements} We would like to thank the following people for 
their contributions and discussions: Michael Carbin, Martin Rinard, Mayank 
Varia, Michael Stone and Somesh Jha. 

\bibliographystyle{abbrv}
\bibliography{proghard,biblio}
\end{document}

%% file: abstract.tex
\begin{abstract}
In this paper, we introduce a formal notion of partial compliance,
called \resistance, of a computer program running together
with a defense mechanism w.r.t a
non-exploitability specification. In our setting, a program may contain
exploitable vulnerabilities, such as buffer overflows, but appropriate defense 
mechanisms built into the program or the operating system render such vulnerabilities hard to exploit by
certain attackers, usually relying on the strength of 
the randomness of a probabilistic transformation of the
environment or the program and some knowledge on the attacker's goals
and attack strategy. We are motivated by the reality that most
large-scale programs have vulnerabilities despite our best efforts to get rid
of them. Security researchers have responded to this state of affairs
by coming up with ingenious defense mechanisms such as address space layout 
randomization (ASLR) or instruction set randomization (ISR) that provide some
protection against exploitation. However, implementations of such mechanism 
have been often shown to be insecure, even against the attacks they were
designed to prevent. By formalizing this notion of attack-resistance we pave 
the way towards addressing the questions:
``How do we formally analyze defense mechanisms? Is there a
mathematical way of distinguishing effective defense mechanisms from
ineffective ones?  Can we quantify and show that these defense
mechanisms provide formal security guarantees, albeit partial, even in
the presence of exploitable vulnerabilities?''. To illustrate our approach
we discuss under which circumstances ISR implementations comply with the \resistance definition.

\end{abstract}